\documentclass[11pt, a4paper]{article}
\usepackage{jheparxiv}
\usepackage[latin1]{inputenc}
\usepackage{amsmath}
\usepackage{amsfonts}
\usepackage{amssymb}
\usepackage{latexsym}
\usepackage{mathrsfs}
\usepackage{graphicx}
\usepackage{color}
\usepackage{slashed}
\usepackage{twistor}
\usepackage[all]{xy}

\renewcommand{\d}{\mathrm{d}}

\subheader{\hfill \texttt{DAMTP-2016-49}}

\title{Twistor methods for AdS\boldmath$_5$}

\author{Tim Adamo, David Skinner and Jack Williams}

\affiliation{Department of Applied Mathematics \& Theoretical Physics \\
        University of Cambridge \\
        Wilberforce Road \\
        Cambridge CB3 0WA, United Kingdom}

\emailAdd{[t.adamo, d.b.skinner, jw729]@damtp.cam.ac.uk}

\abstract{We consider the application of twistor theory to five-dimensional anti-de Sitter space. The twistor space of AdS$_5$ is the same as the ambitwistor space of the four-dimensional conformal boundary; the geometry of this correspondence is reviewed for both the bulk and boundary. A Penrose transform allows us to describe free bulk fields, with or without mass, in terms of data on twistor space. Explicit representatives for the bulk-to-boundary propagators of scalars and spinors are constructed, along with twistor action functionals for the free theories. Evaluating these twistor actions on bulk-to-boundary propagators is shown to produce the correct two-point functions.}

\begin{document}
 
\maketitle

\section{Introduction}

In recent years, twistors have played an important role in studying scattering amplitudes of four-dimensional gauge and gravitational theories. The fundamental tool underlying these investigations is the (linear) Penrose transform~\cite{Penrose:1969ae,Eastwood:1981jy}. This asserts that solutions to massless, free field equations on four-dimensional Minkowski space-time may be described in terms of essentially arbitrary holomorphic functions on twistor space, with the homogeneity of the function determining the helicity of the space-time field. The asymptotic states in scattering processes are taken to obey such free field equations, so twistors are a natural language in which to construct amplitudes.

Twistors also provide a natural arena in which to study four-dimensional CFTs. This is because twistor space carries a natural action of SL(4,$\C$), the (four-fold cover of the) complexification of the space-time conformal group. Here, twistors are closely related to the `embedding space' formalism used in {\it e.g.}~\cite{Dirac:1936fq,Weinberg:2010fx,Penedones:2010ue,Fitzpatrick:2011ia,Paulos:2011ie,SimmonsDuffin:2012uy,Costa:2014kfa} and are particularly useful when considering operators with non-integer spin~\cite{Goldberger:2011yp,Fitzpatrick:2014oza}. In the context of $\cN=4$ SYM, twistor methods have been applied to correlation functions of local gauge invariant operators in {\it e.g.}~\cite{Adamo:2011dq,Adamo:2011cd,Chicherin:2014uca,Koster:2016ebi,Chicherin:2016soh}.

\medskip

By the AdS/CFT correspondence, many four-dimensional CFTs have a dual description as a theory of gravity in five-dimensional anti-de Sitter space~\cite{Maldacena:1997re,Gubser:1998bc,Witten:1998qj}. Given the utility of twistor theory on the boundary side of this correspondence, is natural to ask if it can also be applied in the bulk.  

In this paper, we begin an investigation of the role of twistors in AdS$_5$, following earlier mathematical work in~\cite{Bailey:1998}. After briefly reviewing various descriptions of AdS and its complexification, in section~\ref{geom} we describe its twistor space and the corresponding incidence relations. Remarkably, the {\it twistor} space of AdS$_5$ turns out to be the same as the {\it ambitwistor} space of the boundary space-time. We explore and elucidate this construction in detail. In section~\ref{Penrose} we consider the Penrose transform for free fields on AdS$_5$. Unlike in flat space-time, we show that it is straightforward to describe fields with non-zero mass as well as non-zero spin. From the point of view of AdS/CFT, the most important free fields are bulk-to-boundary propagators and we provide explicit twistor descriptions of these in section~\ref{action}, concentrating on spin-0 and spin-$\half$. We also construct simple twistor actions for these fields and verify that, when evaluated on bulk-to-boundary propagators, they reproduce the expected form for 2-point  correlation functions of boundary operators of the expected conformal weights and spins. We hope that these results will provide a useful starting-point for a twistor reformulation of Witten diagrams.


\section{Geometry}
\label{geom}

The geometry of anti-de Sitter space (or hyperbolic space) is an old and well-studied topic. For the purposes of describing twistor theory in the context of five-dimensional AdS, a particular description of hyperbolic geometry in terms of an open subset of projective space will prove useful. While this description is standard, it is not often utilized in the physics literature so we begin with a brief review of AdS$_5$ geometry from a projective point of view. The twistor space of AdS$_5$ and various aspects of its geometry are then discussed.


\subsection{AdS$_5$ geometry from projective space}

Consider the five-dimensional complex projective space $\CP^5$, charted by homogeneous coordinates encoded in a skew symmetric $4\times4$ matrix $X^{AB}=X^{[AB]}$ with the identification $X\sim \lambda X$ for any $\lambda\in\C^{*}$. For a (holomorphic) metric written in terms of these homogeneous coordinates to be well-defined on $\CP^5$ it must be invariant with respect to the scaling $X\rightarrow \lambda X$ and have no components along this scaling direction (\textit{i.e.}, the metric must not `point off' $\CP^5$ into $\C^6$). The simplest metric satisfying these conditions is
\be\label{metric1}
\d s^{2}=-\frac{\d X^2}{X^2}+\left(\frac{X\cdot\d X}{X^2}\right)^2\,,
\ee
where skew pairs of indices are contracted with the Levi-Civita symbol, $\epsilon_{ABCD}$. This line element is obviously scale invariant, and furthermore has no components in the scale direction. The latter fact follows since the contraction of \eqref{metric1} with the Euler vector field $\Upsilon=X\cdot\frac{\partial}{\partial X}$ vanishes.

Although this metric is projective (in the sense that it lives on $\CP^5$ rather than $\C^6$), it is not global: \eqref{metric1} becomes singular on the quadric
\begin{equation*}
M=\left\{X\in\CP^5 | X^2=0\right\}\subset\CP^5\,.
\end{equation*}
So \eqref{metric1} gives a well-defined metric on the open subset $\CP^{5}\setminus M$. It is a fact that $\CP^5\setminus M$ equipped with this metric is equivalent to complexified AdS$_5$, with the quadric $M$ corresponding to the four-dimensional conformal boundary. Real AdS$_5$, along with a choice of signature (Lorentzian or Euclidean, for instance) is specified by restricting the metric to a particular real slice of $\CP^5$ -- or equivalently, imposing some reality conditions on $X^{AB}$. We will be explicit about these reality conditions below.

To see that \eqref{metric1} really describes AdS$_5$, it suffices to show that it is equivalent to other well-known models of hyperbolic geometry. It is straightforward to see that the metric can be rewritten as
\be\label{metric2}
\d s^2= -\epsilon_{ABCD}\,\d\!\left(\frac{X^{AB}}{|X|}\right)\,\d\!\left(\frac{X^{CD}}{|X|}\right)=-\epsilon_{ABCD}\,\d\mathcal{X}^{AB}\d\mathcal{X}^{CD}\,,
\ee
where $ \mathcal{X}^{AB}:={X^{AB}}/{|X|}$ with $|X| := \sqrt{X^2}$. The coordinates $\mathcal{X}^{AB}$ are invariant under scalings of $X^{AB}$, so they give coordinates on $\C^6$ obeying $\mathcal{X}^2=1$. Since \eqref{metric2} is just the flat metric on $\C^6$, the original metric on $\CP^5\setminus M$ describes a geometry equivalent to the quadric $\mathcal{X}^2=1$ in $\C^6$. With an appropriate choice of reality conditions, this is the well-known model of AdS$_5$ as the hyperboloid in $\R^6$.

To obtain the conformal compactification of AdS$_5$, one includes a conformal boundary isometric to the one-point compactification of 4-dimensional complexified flat space; with appropriate reality conditions this is topologically $S^4$.  We wish to identify this boundary with the quadric $M\subset\CP^5$ on which \eqref{metric1} becomes singular. A point $X\in M$ satisfies $X^2 = 0$ and hence $\det X = 0$.  Since $X^{AB}$ is antisymmetric, non-zero and degenerate, it must have rank 2 and so can be written as the skew of two 4-vectors,
\be\label{bound1}
X^{AB} = C^{[A}D^{B]}\,.
\ee
However, $X$ is projectively invariant under the (separate) transformations
\begin{equation*}
(C,D) \mapsto (C, D + \alpha C)\,, \quad (C,D) \mapsto (C + \beta D, D)\,, \quad (C,D)\mapsto (\gamma C, D)\,, \quad (C,D) \mapsto (C, \delta D)\,
\end{equation*}
for $\alpha,\beta\in\C$, $\gamma,\delta\in\C^{*}$. Performing a sequence of these transformations allows us to assume that $C$ and $D$ take the form
\begin{equation}\label{Plucker}
C = \left(\ \begin{matrix}a\\c\\1\\0\end{matrix}\ \right)
\qquad\text{and}\qquad
D = \left(\ \begin{matrix}b\\d\\0\\1\end{matrix}\ \right),
\end{equation}
where some of $a,b,c$ and $d$ may be infinite, and after which there is no remaining freedom.  Thus, the general form of a boundary point is 
\be\label{bound2}
X^{AB}_{\mathrm{bdry}} = \left( \begin{matrix}\frac{1}{2}x^2 \epsilon^{\dot{\alpha}\dot{\beta}} &&{x^{\dot{\alpha}}}_{\beta}\\-{x_\alpha}^{\dot{\beta}}&&\epsilon_{\alpha\beta}\end{matrix} \right)\,,
\ee
where $\alpha, \dot{\alpha},\ldots$ are dotted and un-dotted two component $\SL(2,\C)$ spinors. The four components of $x^{\alpha\dot\alpha}$ encode the four degrees of freedom in \eqref{Plucker}. Including the point `at infinity,' represented by the \emph{infinity twistor}
\be\label{infinity}
I^{AB}=\left(\begin{matrix} \epsilon^{\dot\alpha\dot\beta} && 0 \\ 0 && 0 \end{matrix}\right)\,,
\ee
gives the one-point compactification of four-dimensional flat space, with $x^{\alpha\dot\beta}$ serving as the usual spinor helicity coordinates. Thus, $M=\{X^2=0\}$ is identified with the $S^4$ conformal boundary of AdS$_5$. The relationship between four-dimensional space-time and simple points in $\CP^5$ is well-established, having appeared in various places in a variety of different guises (\textit{e.g.}, \cite{Dirac:1936fq,Hughston:1984gz,Weinberg:2010fx}).

\medskip

\begin{figure}[t]
\centering
\includegraphics[scale=.3]{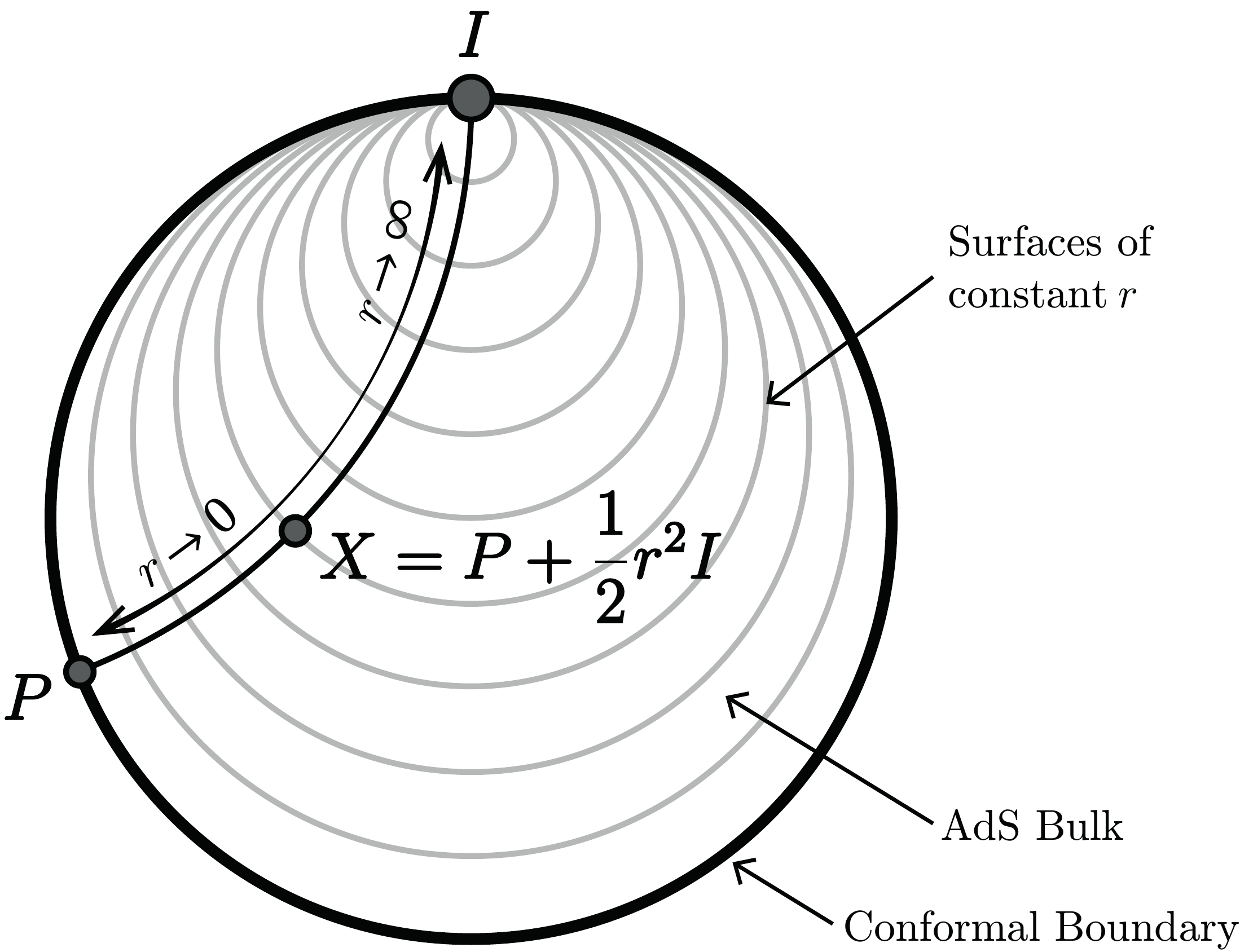}
\caption{Parametrization of AdS space by Poincar\'e coordinates.  The coordinate $r$ controls the distance to the conformal boundary.}
\end{figure}

It is straightforward to obtain other well-known models of AdS$_5$ from the projective one. For example, the Klein model of hyperbolic space is obtained by simply writing the metric \eqref{metric1} using inhomogeneous coordinates on a patch where one of the $X^{AB}$ is non-vanishing. One of the models of AdS used most widely in physical applications is the Poincar\'e model; in Euclidean signature these are global coordinates, and the metric takes the form:
\be\label{Poincare}
\d s^2 =\frac{\d r^2 +\d x_{\alpha\dot\alpha}\d x^{\alpha\dot\alpha}}{r^2}\,,
\ee
with the conformal boundary corresponding to the region where $r\rightarrow 0$.

To obtain Poincar\'e coordinates from the projective model, it suffices to choose a parametrization for $X^{AB}$ in terms of a variable boundary point, $P^{AB}$, of the form \eqref{bound2} and some fixed boundary point. It is convenient to let this fixed boundary point be precisely the infinity twistor \eqref{infinity}, and write:
\be\label{pp1}
X^{AB}=P^{AB}+\frac{r^2}{2} I^{AB}\,.
\ee
As $r \rightarrow 0$, we approach a boundary point $P$, but as $r \to \infty$ with $P$ constant, we approach the fixed infinity twistor. Surfaces of constant $r>0$ correspond to spheres in the bulk of AdS$_5$ which touch the boundary only at $I$.  As $r \rightarrow 0$ this sphere approaches the whole boundary, but as $r \rightarrow \infty$ it shrinks to the single point $I$.  Note that $X^2 = r^2$, so $r$ controls the distance from the conformal boundary. Plugging the parametrization \eqref{pp1} into \eqref{metric1} leads directly to the Poincar\'e metric \eqref{Poincare} after a rescaling of the boundary coordinates $x^{\alpha\dot\alpha}$ by an overall factor of two. This Poincar\'e parametrization will prove useful later when we want to check that certain expressions derived from twistor methods correspond to well-known formulae on space-time.

\medskip

Let us conclude our review of AdS$_5$ geometry with a brief discussion of the reality conditions which can be imposed on the $X^{AB}$ to obtain a real space-time with explicit signature. This is best understood by viewing the metric in terms of the scale-free $\mathcal{X}^{AB}$, constrained to be $\mathcal{X}^2=1$, as in \eqref{metric2}. On $\C^6$, there are two representations of chiral spinors with four components; these are dual to each other, and the bundles of such spinors are denoted by $\mathbb{S}^{A}$, $\mathbb{S}_{A}$ respectively. The coordinates $\mathcal{X}^{AB}$ live in the anti-symmetric square of the first of these: $\mathbb{S}^{A}\wedge\mathbb{S}^{B}$. 

Reality conditions on the $\mathcal{X}^{AB}$ -- and hence the homogeneous coordinates $X^{AB}$ -- correspond to a reality structure on these spinor bundles~\cite{Mason:2011nw,Saemann:2011nb}. Introduce a quaternionic conjugation acting on $Z^{A}\in\mathbb{S}^{A}$ by
\begin{equation*}
Z^{A}=(\mu^{\dot{0}}, \mu^{\dot{1}}, \lambda_{0}, \lambda_{1})\mapsto \hat{Z}^{A} = (-\bar{\mu}^{\dot{1}}, \bar{\mu}^{\dot{0}}, -\bar{\lambda}_{1}, \bar{\lambda}_{0})\,,
\end{equation*} 
which squares to minus the identity: $\hat{\hat{Z}}^{A}=-Z^{A}$. Clearly, there are no real spinors under the $\hat{\cdot}$ -operation, but this conjugation does act involutively on $\mathcal{X}^{AB}$. Restricting to the real slice $\hat{\mathcal{X}}^{AB}=\mathcal{X}^{AB}$ inside $\C^6$ turns \eqref{metric2} into the flat metric on $\R^{1,5}$. This, along with the condition that $\mathcal{X}^2=1$ indicates that these reality conditions describe \emph{Euclidean} AdS$_5$ (the hyperbolic space $\mathbb{H}_5$).

To obtain Lorentzian AdS$_5$ a different reality condition is required. Instead of the quaternionic conjugation, one can take ordinary complex conjugation which exchanges the spinor representations:
\begin{equation*}
Z^{A}\mapsto\overline{Z^{A}}=\bar{Z}_{A}\,.
\end{equation*}
The reality condition on $\mathcal{X}^{AB}$ is then
\begin{equation*}
\overline{\mathcal{X}^{AB}}=\bar{\mathcal{X}}_{AB}=\frac{1}{2}\epsilon_{ABCD}\mathcal{X}^{CD}\,.
\end{equation*}
This real slice results in the flat metric on $\R^{2,4}$, and thus Lorentzian AdS$_5$ as the hyperboloid.


\subsection{The twistor space of AdS$_5$}
\label{TAdS}

It is an interesting fact that the twistor space of AdS$_5$ is the same geometric space as the projective \emph{ambitwistor} space of the complexified, four-dimensional conformal boundary. In any number of dimensions, the projective ambitwistor space of a Riemannian manifold $M_\R$ is the space of complex null geodesics in the complexified manifold $M$~\cite{Witten:1978xx,Isenberg:1978kk,LeBrun:1983,Witten:1985nt}. In the case that $M_\R=S^4$ this ambitwistor space can be written as a quadric in $\CP^{3}\times\CP^3$:
\be\label{ts1}
Q=\left\{(Z^{A},W_{B})\in\CP^{3}\times(\CP^3)^{*} \,|\, Z\cdot W=0\right\}\,,
\ee
where $Z^{A}$, $W_{B}$ are homogeneous coordinates on the two (dual) copies of $\CP^3$, each with its own scaling freedom. The ambitwistor correspondence relates a point in $M$ to a $\CP^{1}\times(\CP^1)^{*}\subset Q$, which can be thought of as the complexified sphere of null directions through that point.

The quadric $Q$ also serves as the \emph{twistor} space of (complexified) AdS$_5$.\footnote{This fact has been known for some time; a mathematical treatment was given by~\cite{Bailey:1998}, and some aspects have also appeared in the physics literature~\cite{Roiban:2000yy,Sinkovics:2004fm,Alday:2010vh}.} The usual twistor correspondence relates a space-time point to an extended geometric object in twistor space, with the intersection theory of these objects encoding the conformal structure of the space-time. To formulate this correspondence, we relate AdS$_5$ to $Q$ by the \emph{incidence relations}:
\be\label{incidence}
Z^{A}=X^{AB}W_{B}\,,
\ee
where $X^{AB}$ describes a point in AdS$_5$. It is easy to see that for a fixed (up to scale) $X$, \eqref{incidence} defines a $\CP^{3}_{X}\subset \CP^3\times(\CP^3)^*$; the fact that $\CP^{3}_{X}\subset Q$ follows from the anti-symmetry of $X^{AB}$ (\textit{i.e.}, the incidence relation preserves $Z\cdot W=0$). Further, since $X^{AB}\in\CP^{5}\setminus M$ it has no kernel so the incidence relation is non-degenerate.

For $Q$ equipped with \eqref{incidence} to be the correct twistor space, the geometry of the incidence relations should capture the conformal geometry of AdS$_5$. To see this, consider two distinct points $X,Y\in\CP^5\setminus M$ and the corresponding $\CP^{3}_{X}$, $\CP^{3}_{Y}\subset Q$. Generically, $\CP^{3}_{X}$ and $\CP^{3}_{Y}$ will intersect in two projective lines in $Q$. To see this, note that $\CP^{3}_{X}\cap\CP^{3}_{Y}$ consists of the points $(Z,W)\in\CP^{3}_{X}$ for which $(X-tY)^{AB}W_{B}=0$ for some $t\in\C^{*}$. The antisymmetric matrix $(X-tY)^{AB}$ has a non-trivial kernel whenever it squares to zero, in which case its kernel is of complex projective dimension one. This shows that $\CP^{3}_{X}\cap\CP^{3}_{Y}$ consists of some number of copies of $\CP^1$. To establish how many, it is useful to write the intersection condition in a scale-free way: 
\begin{equation*}
\left(\frac{X^{AB}}{|X|}-s\frac{Y^{AB}}{|Y|}\right)W_{B}=0 \quad \Leftrightarrow \quad  \left(\frac{X}{|X|}-s\frac{Y}{|Y|}\right)^2=0\,.
\end{equation*}
This gives a quadratic equation in $s$ which has two distinct solutions given by 
\begin{equation}
s_\pm = \frac{X \cdot Y}{|X||Y|} \pm \sqrt{\left(\frac{X \cdot Y}{|X||Y|}\right)^2-1}\,,
\end{equation}
each of which corresponds to an intersection of $\CP^{3}_{X}\cap\CP^{3}_{Y}$ isomorphic to $\CP^1$. 

Generically, these two lines do not themselves intersect because $Y^{AB}$ is non-degenerate. However, when
\be\label{degenerate}
\frac{X}{|X|} \cdot \frac{Y}{|Y|}=1.
\ee
these two solutions degenerate into a single $\CP^1$. Since the geodesic distance $d(X,Y)$ between two points in AdS$_5$ satisfies 
\be\label{geodesic}
\cosh\left(d(X,Y)\right) = \frac{X}{|X|} \cdot \frac{Y}{|Y|},
\ee
the pairs of points satisfying \eqref{degenerate} are precisely those which are null separated. In other words, two points in $\CP^5\setminus M$ are null separated in the AdS conformal structure if and only if their corresponding $\CP^3$s intersect in a single line in twistor space.

A null structure on a (complexified) Lorentzian manifold determines the metric up to a conformal factor.  The null structure given by this degeneracy condition is a canonical choice, so we recover the AdS$_5$ metric \eqref{metric1} up to the conformal factor.  This factor is fixed by making the canonical choice of holomorphic volume form on the $\CP^{3}_{X}$ corresponding to a space-time point:
\be\label{fibreform}
\D^{3} W:=\epsilon^{ABCD}W_{A} \d W_{B}\wedge \d W_{C} \wedge \d W_{D}\,,
\ee
which sets the overall conformal factor in \eqref{metric1} to unity.

\medskip

\begin{figure}[t]
\centering
\includegraphics[scale=.3]{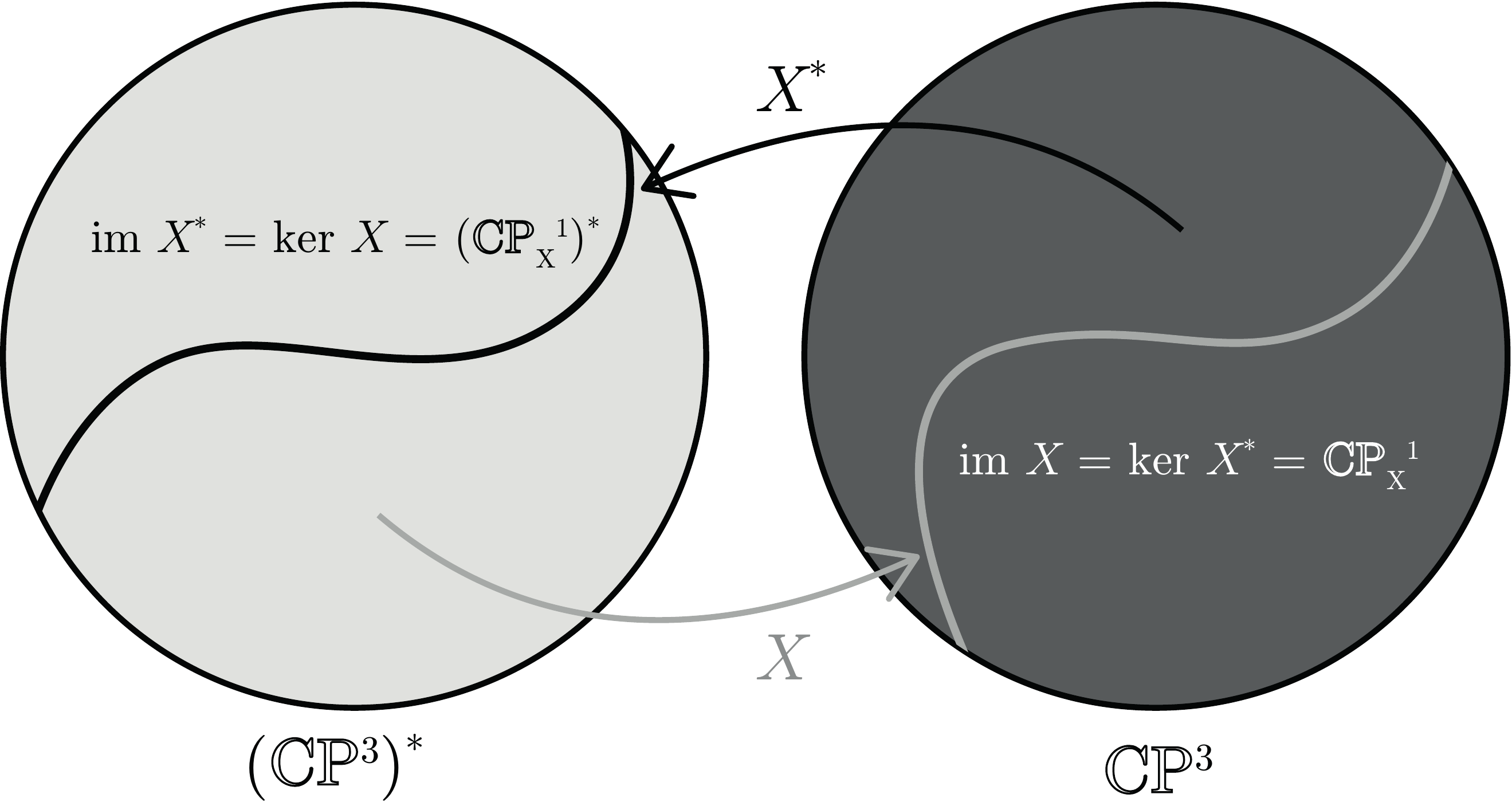}
\caption{Relationship between the linear map $X$ corresponding to a boundary point and its dual.  Both maps determine the canonical $\mathbb{CP}^1 \times \left( \mathbb{CP}^1 \right)^*$ inside $Q$.}
\end{figure}

What happens in twistor space if $X^{AB}$ corresponds to a point on the conformal boundary? This means that $X^2=0$ so $X^{AB}$ has a non-trivial kernel and the incidence relations \eqref{incidence} become degenerate. In particular, since $Z^{A}$ are homogeneous coordinates on $\CP^3$, they cannot all be simultaneously zero -- but there are now solutions of $X^{AB}W_{B}=0$. The space of such solutions has complex projective dimension one, as does the image of $X^{AB}_{\mathrm{bdry}}$ when viewed as a linear map on $(\CP^3)^*$. So for a boundary point $X_{\mathrm{bdry}}$ the degenerate incidence relations are replaced by the linear map
\be\label{bdmap}
X_{\mathrm{bdry}} : \left( \mathbb{CP}^3 \right)^* \setminus(\CP^{1}_{X})^{*}  \rightarrow \CP^{1}_{X}\,,  
\ee
where
\begin{equation*}
(\CP^{1}_{X})^{*}=\left\{X_{\mathrm{bdry}}^{AB}W_{B}=0\right\}\subset (\CP^3)^*\,, \qquad \CP^1_{X}=\left\{X^{\mathrm{bdry}}_{AB}Z^{B}=0\right\}\subset\CP^3\,,
\end{equation*}
are the kernel and image of the linear map, respectively.

In fact, boundary points $X_{\mathrm{bdry}}$ are in one-to-one correspondence with sets $\CP^1_{X} \times (\CP^1_{X})^{*} \subset Q$.  The choice of $\CP^1_{X}\times (\CP^1_{X})^{*}$ determines both the kernel and image of $X^{AB}_{\mathrm{bdry}}$, and any antisymmetric $4 \times 4$ matrix is fixed by these up to an overall scale.  This scale is irrelevant because $X^{AB}$ describes a point in the projective space $\CP^5$. More generally, any subset $\CP^1_{X} \times (\CP^1_{Y})^{*} \subset \CP^{3}\times(\CP^{3})^{*}$ can be specified by two points on the conformal boundary $x^{\alpha\dot\alpha}$, $y^{\alpha\dot\alpha}$ as in \eqref{bound2}. The condition that this subset lies inside $Q$, namely that $Z\cdot W=0$ imposes the constraint $x^{\alpha\dot\alpha}=y^{\alpha\dot\alpha}$. Hence, the two lines $\CP^1_{X} \times (\CP^1_{Y})^{*} \subset Q$ correspond to the \emph{same} point on the four-dimensional boundary.

This establishes the geometry of twistor space for both the bulk and boundary of AdS$_5$. A point in the bulk corresponds to a $\CP^3$ inside $Q$; for boundary points this correspondence degenerates to give the standard ambitwistor relation between a point on the boundary and a $\CP^1\times(\CP^1)^*$ inside $Q$.

\medskip

\begin{figure}[t]
\centering
\includegraphics[scale=.3]{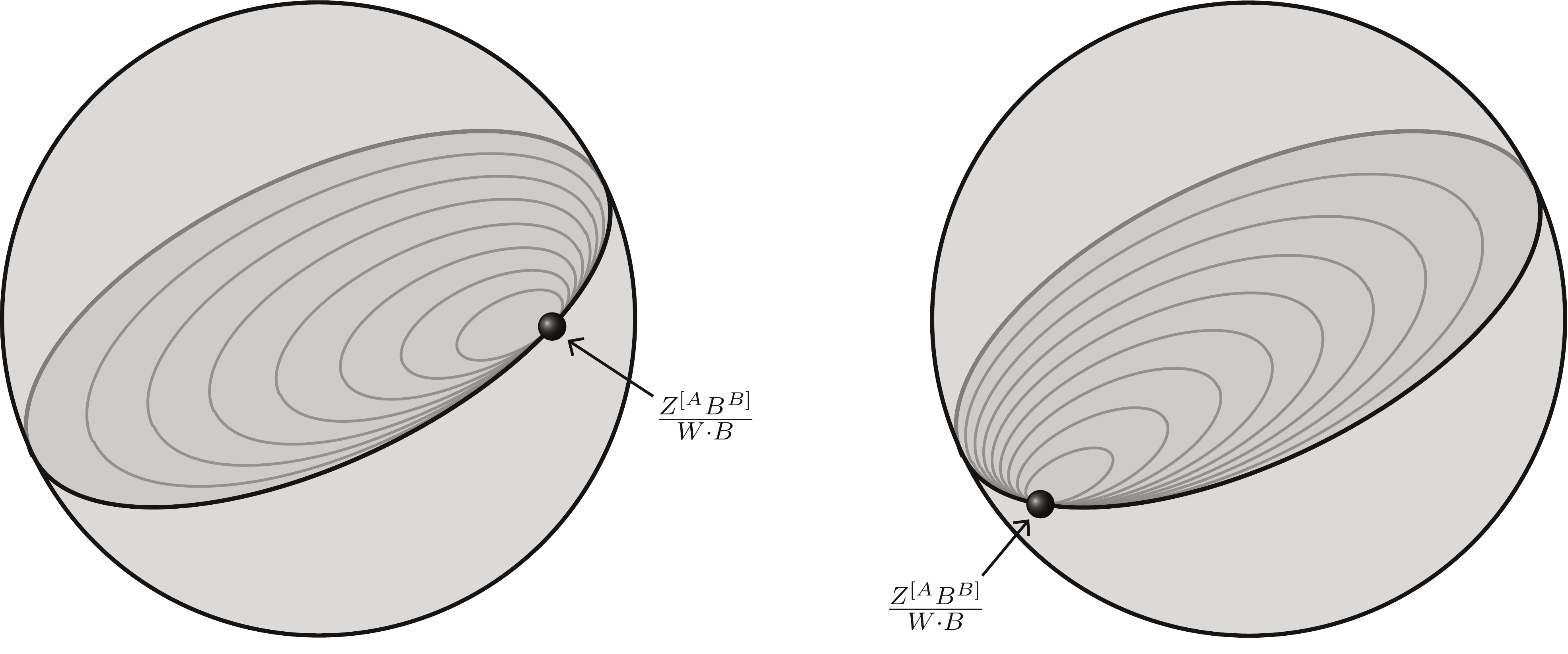}
\caption{The totally null set of points in spacetime $\mathbb{CP}^5$ corresponding to a twistor point $(Z,W)$ for two different choices of $B$.  We view $B$ as fixed and vary $A$, tracing out a three-dimensional space of solutions.  Changing $B$ alters the parametrization of this solution space, but not the set itself.}
\end{figure}

It is equally natural to ask for the twistor correspondence in the other direction: what does a point in twistor space correspond to in space-time? Given fixed $(Z,W) \in Q$, we want to know which space-time points $X$ satisfy the incidence relations
\begin{equation*} 
Z^A = X^{AB}W_B\,.
\end{equation*}
The solution set consists of points of the form
\begin{equation}
X^{AB} = \frac{Z^{[A}B^{B]}}{W \cdot B} + \epsilon^{ABCD} A_C W_D,
\end{equation}
where $A_C$ is an arbitrary parameter and $B^B$ is an arbitrary twistor with $B \cdot W \neq 0$.  Transformations of the form $A \mapsto A + \alpha W$ leave $X$ invariant so the space of solutions is three-dimensional.  Making a different choice of $B$ can be accommodated by a redefinition of $A$, so $B$ contributes no further degrees of freedom.  Moreover, any tangent vector to this set is a null vector of the form $\epsilon^{ABCD} (\delta A)_C W_D$, where $\delta A$ is a displacement in the parameter $A$. Thus, a point in twistor space corresponds to a totally null three-plane in $\CP^5$ and hence AdS$_5$.

\begin{figure}[t]
\centering
\includegraphics[scale=.4]{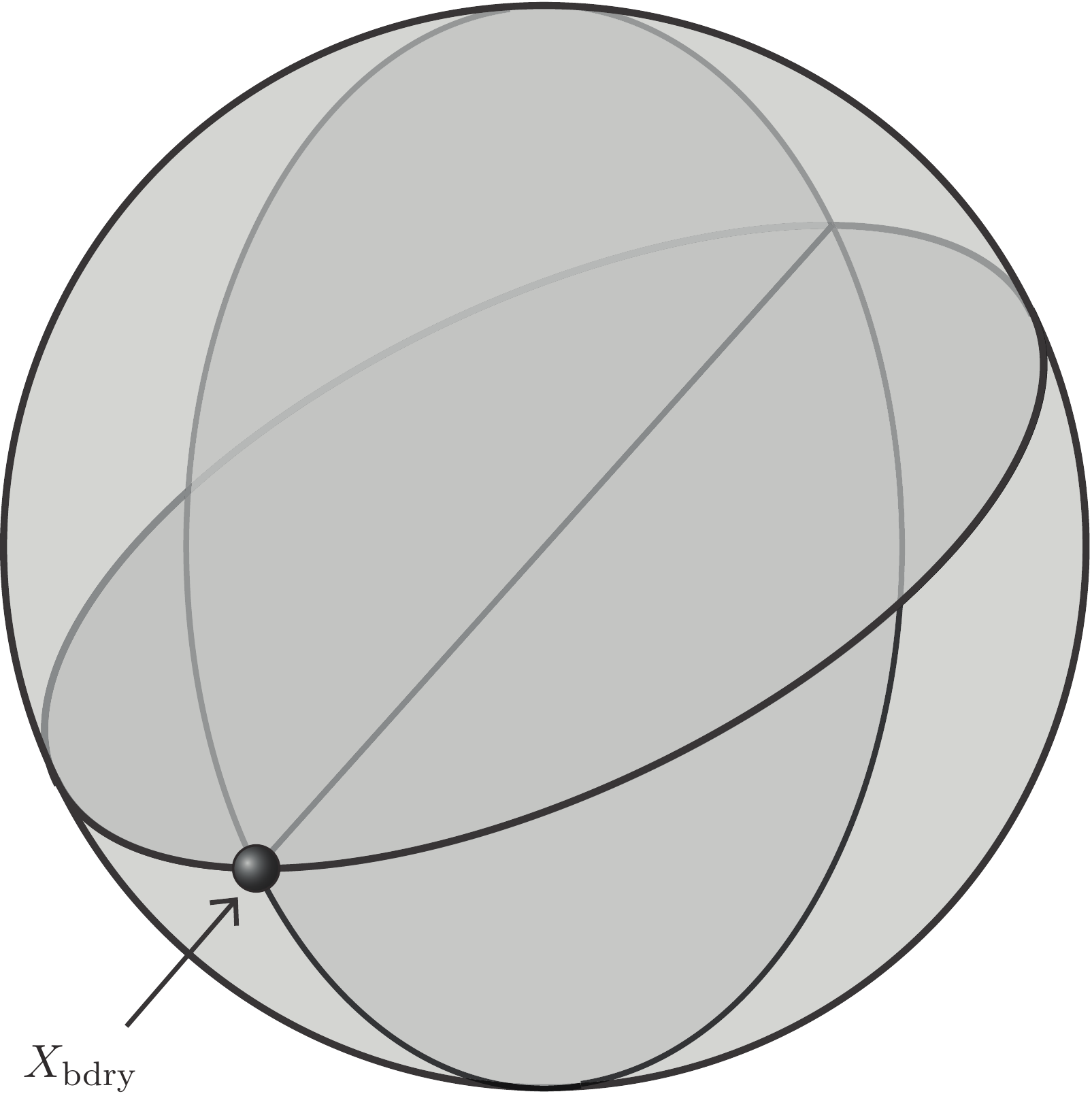}
\caption{The set of points in spacetime $\mathbb{CP}^5$ corresponding to two different twistor points $(Z,W)$ and $(\tilde{Z},\tilde{W})$.  The intersection is generically one-dimensional, but if $Z \cdot \tilde{W} = 0$ and $\tilde{Z} \cdot W = 0$, then it is two dimensional and the closure includes the boundary point whose canonical $\mathbb{CP}^1_X \times \left( \mathbb{CP}^1_X \right)^* \subset Q$ contains $(Z,W)$ and $(\tilde{Z},\tilde{W})$.}
\end{figure}

How can two such three-planes intersect? Let $(Z,W),(\tilde{Z},\tilde{W})\in Q$ be distinct twistor points; the general projective solution to the simultaneous equations
\begin{align}
\begin{matrix}
Z^A &= X^{AB}W_B, \\
\widetilde{Z}^A &= X^{AB}\widetilde{W}_B
\end{matrix}
\end{align}
is
\begin{align}
X^{AB} = \alpha& \left( Z^{[A}B^{B]} \frac{\widetilde{Z} \cdot W}{W \cdot B} - \widetilde{Z}^{[A}\widetilde{B}^{B]} \frac{Z \cdot \widetilde{W}}{\widetilde{W} \cdot \widetilde{B}} + B^{[A}\widetilde{B}^{B]} \frac{(Z \cdot \widetilde{W})(\widetilde{Z} \cdot W)}{(B \cdot W)(\widetilde{B} \cdot \widetilde{W})} \right) \nonumber\\
& \qquad + \gamma\, \epsilon^{ABCD} \,W_C\, \widetilde{W}_D,
\end{align}
where $\alpha \neq 0$ and $B,\tilde{B}$ are such that $B \cdot \tilde{W} = 0$ and $\tilde{B} \cdot W = 0$ while $B \cdot W \neq 0$, $\tilde{B} \cdot \tilde{W} \neq 0$.   This solution is parametrized by the two complex numbers $\alpha,\gamma$, or equivalently, a projective line. So for all pairs $(Z,W)$ and $(\tilde{Z},\tilde{W})$ the corresponding null three-planes intersect in a line in $\CP^5$. 

However, if $Z \cdot \tilde{W} = 0$ and $\tilde{Z} \cdot W = 0$, then further solutions are possible.  In this case, the general solution is
\begin{align}\label{intersection}
X^{AB} = \alpha \frac{Z^{[A}B^{B]}}{W \cdot B} + \beta \frac{\widetilde{Z}^{[A}\widetilde{B}^{B]}}{\widetilde{W} \cdot \widetilde{B}} + \gamma\, \epsilon^{ABCD}\, W_C\, \widetilde{W}_D,
\end{align}
with $\alpha,\beta \neq 0$.  This gives a two-dimensional projective space of solutions, parametrized by homogeneous coordinates $(\alpha,\beta,\gamma)$. Note that the conditions $Z \cdot \tilde{W} = 0$ and $\tilde{Z} \cdot W = 0$ mean that $(Z,W)$ and $(\tilde{Z},\tilde{W})$ lie inside $\mathbb{CP}^1_X \times \left( \mathbb{CP}^1_X \right)^* \subset Q$ for a boundary point 
\begin{equation*}
X^{AB}_{\text{bdry}} = \epsilon^{ABCD}\, W_C\, \widetilde{W}_D\,.
\end{equation*}
This point is in the closure of the two-dimensional intersection of their three-planes \eqref{intersection} but not in the solution space itself, since it requires $(\alpha,\beta,\gamma)=(0,0,1)$.\footnote{It is interesting to contrast this against the situation for the twistor space of $\C^6$. There twistor points define totally null 3-planes which do not intersect generically, and only intersect in a line if their twistor points obey a nullity relation akin to $Z\cdot\tilde{W}+\tilde{Z}\cdot W=0$ ~\cite{Hughston:1986hb}.}


\section{The Penrose transform}
\label{Penrose}

A basic property of twistor theory in any number of dimensions is its ability to encode fields living on space-time in terms of geometric data on twistor space. In four space-time dimensions the basic tool in this regard is the \emph{Penrose transform}, relating solutions of the zero-rest-mass equations to certain cohomology classes on twistor space~\cite{Penrose:1969ae,Eastwood:1981jy}. It is less widely known that the Penrose transform extends to any number of space-time dimensions, where cohomology of the corresponding twistor space encodes solutions to certain equations on space-time~\cite{Baston:1989}.

We want to describe fields on AdS$_5$ in terms of some geometric data on the twistor space $Q$. Simple examples of such fields are massive scalars or spinors, which obey field equations
\be\label{mfs}
\Box_{\mathrm{AdS}}\Phi - m^2\, \Phi =0\,, \qquad \slashed{\D}_{\mathrm{AdS}} \Psi = m\,\Psi\,,
\ee
respectively, with $\Box_{\mathrm{AdS}}$ the AdS$_5$ Laplacian and $\slashed{\D}_{\mathrm{AdS}}$ the AdS$_5$ Dirac operator. For such scalar and spinor fields in AdS$_5$ it is well-known that their masses obey relations:
\be\label{smass}
m^2=\Delta (\Delta-4)\,,
\ee
for the scalar, and 
\be\label{fmass}
|m|=\Delta-2\,,
\ee
for the spinor. The parameter $\Delta$ controls the asymptotic behaviour of the fields near the AdS boundary, and is also the conformal dimension of the local operator in the boundary CFT$_4$~\cite{Witten:1998qj}.


\subsection{Scalars: Direct and indirect transform}

Functions of specific homogeneity in $Z$ and $W$ form a natural set of line bundles on $Q$. In particular, denote the line bundle of holomorphic functions scaling as
\begin{equation*}
f(\alpha Z, \beta W) = \alpha^{m}\,\beta^{n}\,f(Z,W)\,, \qquad \alpha,\beta\in\C^{*}\,,
\end{equation*}
by $\cO(m,n)\rightarrow Q$. The line bundles $\cO(m,n)$ can be tensored with other bundles over $Q$ to form weighted bundles of geometric objects with the specified scaling properties.

For some fixed scaling dimension $\Delta$, consider a $(0,3)$-form on $Q$ taking values in $\cO(-\Delta,\Delta-4)$, denoted by $f\in\Omega^{0,3}(Q,\cO(-\Delta,\Delta-4))$. The bundle $\cO(-\Delta,\Delta-4)$ is only well-defined if $\Delta\in\Z$, but this is consistent with the expected conformal dimensions of boundary operators dual to bulk scalars. Restricting $f$ to the $\CP^3_{X}\subset Q$ corresponding to the AdS$_5$ point $X$ is accomplished simply by imposing the incidence relations:
\begin{equation*}
f(Z^{A},W_{B})|_{X}=f(X^{AC}W_C , W_B)\,.
\end{equation*}
So $f|_X$ is a $(0,3)$-form on $\CP^5\times\CP^{3}_{X}$ which is homogeneous of degree $-\Delta$ in $X$ and $-4$ in $W$. Integrating $f|_X$ over $\CP^{3}_{X}$, we define
\be\label{dpt1}
\Phi(X)=|X|^{\Delta} \int\limits_{\CP^{3}_{X}} \D^{3}W\wedge f|_{X}\,.
\ee
Clearly, $\Phi$ is homogeneous of degree zero in $X$ (\textit{i.e.}, $X\cdot\partial \Phi=0$), and hence a well-defined scalar field on AdS$_5$ rather than a section of some line bundle over $\CP^5$. Further, it is an easy consequence of the incidence relations that $\Phi$ obeys
\begin{equation*}
\frac{\partial}{\partial X}\cdot \frac{\partial}{\partial X} \left(|X|^{-\Delta} \Phi\right) = 0\,,
\end{equation*}
if and only if $f$ is holomorphic, $\dbar f = 0$. Since any $f$ which is $\dbar$-exact integrates to zero, we see that $\Phi(X)$ is determined by the cohomology class $[f]\in H^{0,3}(Q,\cO(-\Delta,\Delta-4))$ on twistor space. 

A straightforward calculation reveals that
\begin{equation*}
X\cdot\frac{\partial}{\partial X} \Phi=0=\frac{\partial}{\partial X}\cdot \frac{\partial}{\partial X} \left(|X|^{-\Delta} \Phi\right) \quad \Leftrightarrow \quad \Box_{\mathrm{AdS}}\Phi=\Delta(\Delta-4)\Phi\,.
\end{equation*}
Thus, any $f$ which is a cohomology class defines a solution to the scalar equation of motion with appropriate scaling dimension $\Delta$ via the integral construction \eqref{dpt1}. An argument in homological algebra can be used to show that in fact \emph{every} massive scalar on AdS$_5$ -- subject to suitable analyticity conditions -- can be represented in this way~\cite{Baston:1989,Bailey:1998}. We refer to this correspondence as the \emph{direct Penrose transform}:
\be\label{dpt2}
H^{0,3}\!\left(Q,\cO(-\Delta,\Delta-4)\right)\cong \left\{ \Phi(X) \mbox{ on AdS}_5\, |\, \Box_{\mathrm{AdS}}\Phi=\Delta(\Delta-4)\Phi\right\}\,,
\ee
the isomorphism being realized from left to right by the integral formula \eqref{dpt1}.

\medskip

Unlike in four-dimensions, the Penrose transform in $d>4$ is not unique. For AdS$_5$, this non-uniqueness takes two different forms. The first of these is rather trivial, following from the fact that for bulk points $X$, the incidence relations can be inverted:
\be\label{inverseir}
Z^{A}=X^{AB}W_{B} \Leftrightarrow W_{B}=\frac{X_{BC}}{X^2}Z^{C}\,.
\ee
These inverted relations associate a `dual' $\CP^3$ to $X$ which is now parametrized by $Z$ rather than $W$; we denote this dual by $(\CP^{3}_{X})^{*}$. Interchanging homogeneities of $Z$ and $W$ for a cohomology class then gives an alternative representation for any scalar of scaling dimension $\Delta$ via
\be\label{dpt3}
\Phi(X)=|X|^{-\Delta}\!\! \int\limits_{(\CP^{3}_{X})^\vee} \D^{3}Z\wedge \tilde{f}|_{X}\,, \qquad \tilde{f}\in H^{0,3}\!\left(Q,\cO(\Delta-4,-\Delta)\right)\,.
\ee

More non-trivial is the \emph{indirect Penrose transform}, which describes AdS scalars by elements of an entirely different cohomology group:
\be\label{ipt1}
H^{0,2}\!\left(Q,\cO(1-\Delta,\Delta-3)\right)\cong \left\{ \Phi(X) \mbox{ on AdS}_5\, |\, \Box_{\mathrm{AdS}}\Phi=\Delta(\Delta-4)\Phi\right\}\,.
\ee
The existence of this alternative description for a scalar $\Phi$ is related to a certain obstruction problem in twistor space~\cite{Baston:1989,Mason:2011nw,Saemann:2011nb}. In particular, allowing $g$ to extend off the quadric $Z\cdot W=0$ in $\CP^3\times(\CP^3)^*$ relates $g$ to a direct Penrose transform representative by
\be\label{ipt2}
\dbar g= (Z\cdot W)\, f\,, 
\ee
where $f$ takes values in $H^{0,3}(Q,\cO(-\Delta,\Delta-4))$. 

Equation \eqref{ipt2} can be used to produce an integral formula for $\Phi$ in terms of $g$ (this is adapted from a similar argument for the indirect transform for flat 6-dimensional space-time~\cite{Mason:2011nw}). Extending off the quadric is accomplished at the level of the incidence relations by imposing
\be\label{iptir}
Z^{A}=X^{AB}\left(W_{B}+\delta W_{B}\right)\,,
\ee
for some `small' $\delta W_{B}$ and then considering the limit as $\delta W_{B}\rightarrow 0$. The space-time scalar is then defined in terms of $g$ by:
\be\label{ipt3}
\Phi(X)=|X|^{\Delta}\lim_{\delta W\rightarrow 0} \int\limits_{\CP^{3}_{X}}\D^{3}W\wedge\left.\left(\frac{\dbar g}{Z\cdot W}\right)\right|_{Z^{A}=X^{AB}(W_{B}+\delta W_{B})}\,.
\ee
Note that `dual' representatives for the indirect Penrose transform are also constructed using the inverted incidence relations \eqref{inverseir}; this amounts to describing $\Phi$ by $\tilde{g}\in H^{0,2}(Q,\cO(\Delta-3,1-\Delta))$ in the obvious way.


\subsection{Spinors: direct and indirect transform}

A generic eight-component spinor field on AdS$_5$ can be separated into a chiral and anti-chiral parts, taking values in $\mathbb{S}_{A}$ or $\mathbb{S}^{A}$, respectively. Without loss of generality, consider those components with a downstairs spinor index, of the form $\Psi_{A}(X)$. The equation of motion for a chiral spinor in the projective description of AdS$_5$ is:
\be\label{seom1}
\left(\slashed{\D}\Psi\right)^{B} = \left(\Delta-2\right)\frac{X^{AB}}{|X|}\Psi_{A}\,,
\ee
where the relation $|m|=\Delta-2$ has been used. This equation is further simplified upon noting that the Dirac operator acts as
\begin{equation*}
 \left(\slashed{\D}\Psi\right)^{B}=|X|\frac{\partial}{\partial X_{AB}} \Psi_{A} - 2 \frac{X^{AB}}{|X|}\Psi_{A}\,,
\end{equation*}
to leave
\be\label{seom2}
|X|\partial^{AB}\Psi_{A}=\Delta \frac{X^{AB}}{|X|}\Psi_{A}\,.
\ee

On the twistor space $Q$, these fields are described by a $(0,3)$-form with values in $\cO(-\Delta-\frac{1}{2},\Delta-\frac{9}{2})$, for fixed $\Delta$. This bundle is only well defined for $\Delta\in\Z +\frac{1}{2}$, which is again consistent with the expected conformal dimensions of spinor primary operators on the boundary. For $\psi\in \Omega^{0,3}(Q,\cO(-\Delta-\frac{1}{2},\Delta-\frac{9}{2}))$ we  form a space-time spinor field as
\be\label{sdpt1}
\Psi_{A}(X)=|X|^{\Delta+\frac{1}{2}}\int\limits_{\CP^{3}_X} \D^{3}W\wedge W_{A}\,\psi|_{X}\,,
\ee
where again $\psi|_{X}$ denotes that the incidence relations have been imposed. It is straightforward to show that the equation of motion \eqref{seom2} holds for $\Psi_A$ if and only if $\dbar\psi=0$. Once more, a homological argument demonstrates that every chiral, massive spinor on AdS$_5$ can be represented by \eqref{sdpt1} for some choice of $\psi$ in the relevant cohomology~\cite{Baston:1989,Bailey:1998}. This gives the \emph{direct} Penrose transform for spinors:
\be\label{sdpt2}
H^{0,3}\!\left(Q,\cO\left(-\Delta-\frac{1}{2},\Delta-\frac{9}{2}\right)\right)\cong \left\{ \Psi_{A}(X) \mbox{ on AdS}_5\, |\, (\slashed{\D}\Psi)^{B}=\Delta \frac{X^{AB}}{|X|} \Psi_{A}\right\}\,.
\ee
The integral formula \eqref{sdpt1} realizes this isomorphism from the left to the right.

Just like in the case of the scalar, there is an \emph{indirect} version of the Penrose transform for spinors, given by
\be\label{sipt1}
H^{0,2}\!\left(Q,\cO\left(\frac{3}{2}-\Delta,\Delta-\frac{5}{2}\right)\right)\cong \left\{ \Psi_{A}(X) \mbox{ on AdS}_5\, |\, (\slashed{\D}\Psi)^{B}=\Delta \frac{X^{AB}}{|X|} \Psi_{A}\right\}\,.
\ee
The existence of an indirect transform is again related to an obstruction problem in twistor space~\cite{Baston:1989,Mason:2011nw}, with any indirect representative $\chi$ related to a direct representative $\psi$ by
\be\label{sipt2}
\dbar\chi=(Z\cdot W)^{2}\,\psi\,.
\ee
Using this, an integral formula for the indirect transform is given by extending off the quadric in a similar fashion to the scalar:
\be\label{sipt3}
\Psi_{A}(X)=|X|^{\Delta+\frac{1}{2}} \lim_{\delta W\rightarrow 0} \int\limits_{\CP^{3}_{X}}\D^{3}W\wedge W_{A} \left.\left(\frac{\dbar \chi}{(Z\cdot W)^2}\right)\right|_{Z^{A}=X^{AB}(W_{B}+\delta W_{B})}\,.
\ee
Note that there are `dual' versions of both the direct and indirect transform; both are given by swapping the weights of $Z$ and $W$, corresponding to the inverse incidence relations \eqref{inverseir}.


\section{Free theory, Bulk-to-boundary propagators \& 2-point functions}
\label{action}

In applications of twistor theory to Minkowski space, the Penrose transform can be used to encode physically relevant external states in terms of twistor data. A basic example relevant for scattering amplitude calculations is a momentum eigenstate: an on-shell space-time field modelled on $\e^{ik\cdot x}$ is encoded in terms of certain distributional cohomology classes on twistor space~\cite{Adamo:2011pv,Mason:2011nw}. In AdS, the S-matrix is replaced by correlation functions of specified boundary data for the space-time fields~\cite{Gubser:1998bc,Witten:1998qj}. In this setup the appropriate external states are bulk-to-boundary propagators that propagate the boundary data into the AdS bulk. Computing the tree-level $n$-point correlation functions in the bulk boils down to extracting that piece of the classical generating functional which is multilinear in these external states on the AdS background.

In this section, we demonstrate that the most basic part of this AdS/CFT dictionary can be translated to twistor space by giving explicit representatives for scalar and spinor bulk-to-boundary propagators. The two-point functions for these fields are then derived in a purely twistorial manner by writing the free bulk theory in twistor variables.


\subsection{Scalars}

Holomorphic, first-order action functionals present a natural candidate for describing free theories on twistor space. For direct representatives, such an action is simply:
\be\label{hCS1}
S[f,h]=\int \D^{3}Z\wedge\D^{3}W\wedge\bar{\delta}(Z\cdot W)\wedge h\wedge\dbar f\,,
\ee
where the top-degree holomorphic form on $Q$ is written as
\begin{equation*}
\int \D^{3}Z\wedge\D^{3}W\wedge\bar{\delta}(Z\cdot W) = \oint \frac{\D^{3}Z\wedge\D^{3}W}{Z\cdot W}\,,
\end{equation*}
with the holomorphic delta function $\bar{\delta}(Z\cdot W)$ equivalent to a contour integral localizing the measure to the quadric $Z\cdot W=0$ inside $\CP^3\times\CP^3$. This measure is a $(5,0)$-form on $Q$ valued in $\cO(3,3)$. 

This action is a functional of $f\in\Omega^{0,3}(Q,\cO(-\Delta,\Delta-4))$ and $h\in\Omega^{0,1}(Q,\cO(\Delta-3, 1-\Delta))$ and its field equations are simply
\begin{equation*}
\dbar f = 0 = \dbar h\,,
\end{equation*}
imposing that $f$ and $h$ are cohomology classes on-shell. By \eqref{dpt2}, it follows that $\dbar f=0$ corresponds to the equation of motion $\Box_{\mathrm{AdS}}\Phi=\Delta(\Delta-4)\Phi$ for a scalar field. The second field equation, $\dbar h=0$, is actually non-dynamical on space-time, as the cohomology group $H^{0,1}(Q, \cO(\Delta-3,1-\Delta))$ is empty~\cite{Baston:1989,Bailey:1998}.  Hence, $h$ is just a Lagrange multiplier and solutions to the field equations are in one-to-one correspondence with solutions to the massive scalar equation of motion on AdS$_5$.

It is easy to see that the action \eqref{hCS1} is not suitable for computing any observables in the bulk theory, though. Indeed, the action vanishes when evaluated on solutions to the equations of motion, whereas the appropriate space-time action is equal to a boundary term when evaluated on extrema. So although \eqref{hCS1} gives the correct equations of motion, it is \emph{not} equivalent to the free space-time action. This is analogous to the difference between space-time actions with kinetic terms $\partial \Phi \cdot \partial\Phi$ and $\Phi \Box\Phi$: they have the same equations of motion, although the former is equal to a boundary term on-shell whereas the latter vanishes.

To write a twistor action with non-vanishing extrema, the variational problem must involve an indirect representative $g\in\Omega^{0,2}(Q,\cO(1-\Delta,\Delta-3))$ and its dual $\tilde{g}\in\Omega^{0,2}(Q,\cO(\Delta-3,1-\Delta))$ coupled to fixed `sources' in the twistor space. For a given $\Delta$ these sources are specified by a cohomology class $f\in H^{0,3}(Q,\cO(-\Delta, \Delta-4))$ and its dual $\tilde{f}\in H^{0,3}(Q,\cO(\Delta-4,-\Delta))$ which are \emph{not} part of the variational problem. The action is:
\begin{equation}\label{sta1}
 S[g,\tilde{g}]=\int\D^{3}Z\wedge\D^{3}W\wedge\left[\bar{\delta}^{\prime}(Z\cdot W)\wedge\tilde{g}\wedge\dbar g - \bar{\delta}(Z\cdot W)\wedge f\wedge\tilde{g} + \bar{\delta}(Z\cdot W)\wedge\tilde{f}\wedge g\right]\,, 
\end{equation}
where $\bar{\delta}^{\prime}(Z\cdot W)=\dbar(Z\cdot W)^{-2}$ is the $(0,1)$-distribution which acts like a derivative of a delta function. Since $f,\tilde{f}$ themselves constitute direct Penrose transform representatives, this is not the usual picture one has for physical sources. Instead, one should view $f,\tilde{f}$ as arising from an auxiliary variational problem, akin to the action \eqref{hCS1}.

The equations of motion arising from \eqref{sta1} are
\be\label{steom1}
\bar{\delta}^{\prime}(Z\cdot W)\,\dbar g=\bar{\delta}(Z\cdot W)\, f \quad \Leftrightarrow \quad \dbar g = (Z\cdot W)\, f\,,
\ee
\begin{equation*}
\bar{\delta}^{\prime}(Z\cdot W)\,\dbar \tilde{g}=\bar{\delta}(Z\cdot W)\, \tilde{f} \quad \Leftrightarrow \quad \dbar \tilde{g} = (Z\cdot W)\, \tilde{f}\,,
\end{equation*}
which are precisely the correct on-shell conditions \eqref{ipt2} for indirect Penrose transform representatives. This refined action is non-vanishing when evaluated on solutions to these equations of motion:
\be\label{sta2}
\left.S[g,\tilde{g}]\right|_{\mathrm{on-shell}}=\int\D^{3}Z\wedge\D^{3}W\wedge\bar{\delta}^{\prime}(Z\cdot W)\wedge\tilde{g}\wedge\dbar g\,.
\ee
So although \eqref{sta1} requires the addition of source terms, it leads to sensible equations of motion and is non-zero when evaluated on extrema, making it a good candidate for computing AdS$_5$ observables in twistor space. The two-point function should be given by \eqref{sta2}, where $g,\tilde{g}$ are chosen to represent the external states: bulk-to-boundary propagators.

\medskip

For a massive scalar on AdS$_5$, the bulk-to-boundary propagator $K_\Delta$ is a solution to the equation of motion which becomes proportional to a delta function on the boundary. In Poincar\'{e} coordinates, these conditions read:
\begin{equation*}
\Box_{\mathrm{AdS}}K_{\Delta}(r, x;y)=\Delta(\Delta-4)\,K_{\Delta}(r, x; y)\,, \qquad \lim_{r\rightarrow 0} r^{\Delta-4} K_{\Delta}(r, x;y)=\delta^{4}(x-y)\,,
\end{equation*}
where $(r,x^{\alpha\dot\alpha})$ is a bulk point in AdS$_5$, and $y^{\alpha\dot\alpha}$ is a point on the boundary $S^4$. An expression for this bulk-to-boundary propagator is given in Poincar\'e coordinates by
\be\label{btb1}
 K_{\Delta}(r, x;y)= \mathrm{c}_{\Delta} \left(\frac{r}{r^2 +(x-y)^2}\right)^{\Delta}\,, 
\ee
where $\mathrm{c}_{\Delta}$ is an overall normalization which will be ignored from now on.

How is \eqref{btb1} presented on twistor space? Since $K_{\Delta}$ is a solution to the equation of motion, it should be representable by the Penrose transform. Consider the distributional form
\be\label{btbd}
f_{\Delta}(Z,W)=[AB]^{\Delta}\,\frac{\bar{\delta}^{3}_{\Delta-4}(W,A)}{(Z\cdot B)^{\Delta}} \,,
\ee
where $A_{A},B_{A}$ are two fixed points in $\CP^{3}$, and $[AB]=I^{CD}A_{C}B_{D}$ denotes the contraction of $A$ and $B$ with the infinity twistor of the boundary. The delta function $\bar{\delta}^{3}_{\Delta-4}(W,A)$ is defined as
\begin{equation*}
 \bar{\delta}^{3}_{\Delta-4}(W,A)=\int \frac{\d t}{t}\, t^{\Delta}\,\bigwedge_{A=1}^{4}\dbar \left(\frac{1}{W_{A}+t\,A_{A}}\right)\,.
\end{equation*}
This gives a $(0,3)$-form distribution enforcing the projective coincidence of its two arguments which is homogeneous of degree $\Delta-4$ in $W$ and $-\Delta$ in $A$. 

Up to singularities determined entirely by the fixed points $A,B$, this object is $\dbar$-closed and is homogeneous of degree zero in $A,B$. Thus, \eqref{btbd} can be treated as a class in $H^{0,3}(Q,\cO(-\Delta,\Delta-4))$, so the direct Penrose transform can be applied to give a space-time field
\begin{equation*}
|X|^{\Delta}\int\limits_{\CP^{3}_X}\D^{3}W\wedge\frac{\bar{\delta}^{3}_{\Delta-4}(W,A)}{(X^{CD}W_{D}B_{C})^{\Delta}}[AB]^{\Delta} = \frac{|X|^{\Delta} [AB]^{\Delta}}{(X^{CD}A_{C}B_{D})^{\Delta}}\,.
\end{equation*}
Notice that $A$ and $B$ only appear as the skew-symmetric combination $Y_{CD}=A_{[C}B_{D]}$ through $[AB]$, $X^{CD} A_{C}B_{D}$ in the final answer. Since $Y^2=0$, this corresponds to a fixed point on the boundary of AdS$_5$, so:
\be\label{btbd2}
\int\limits_{\CP^3_X}\D^{3}W\wedge f_{\Delta}|_{X}=\frac{|X|^{\Delta} (I\cdot Y)^{\Delta}}{(X\cdot Y)^{\Delta}}\,.
\ee
It is easy to confirm (by going to the Poincar\'e parametrization, for instance) that this expression is equal to \eqref{btb1}. Thus, \eqref{btbd} is a direct transform representative for the scalar bulk-to-boundary propagator.

An indirect representative for the bulk-to-boundary propagator is given by
\be\label{btbi}
g_{\Delta}(Z,W)=[AB]^{\Delta}\int s^{\Delta-1}\d s\,\frac{\bar{\delta}^{3}_{\Delta-3}(W,A(s))}{(Z\cdot A)^{\Delta-1}}\,,
\ee
where $A(s)=A+sB$ parametrizes a point on the projective line spanned by $A\wedge B$ in $\CP^3$. The integral over the parameter $s$ reduces the distributional form degree of $g$ to $(0,2)$, and it is easy to show that \eqref{btbi} is homogeneous of degree $1-\Delta$ in $Z$ and $\Delta-3$ in $W$. Note that $g$ is not obviously $\dbar$-closed, as
\be\label{sid1}
\dbar g_{\Delta}=[AB]^{\Delta}\int s^{\Delta-1}\d s\,\bar{\delta}^{(\Delta-2)}(Z\cdot A)\,\bar{\delta}^{3}_{\Delta-3}(W,A(s))\,,
\ee
where $\bar{\delta}^{(\Delta-2)}(Z\cdot A)$ is a $(0,1)$-distribution acting like the $(\Delta-2)^{\mbox{th}}$-derivative of a delta-function:
\begin{equation*}
 \bar{\delta}^{(\Delta-2)}(Z\cdot A):=\dbar\left(\frac{1}{(Z\cdot A)^{\Delta-1}}\right)\,.
\end{equation*}
However, by integrating \eqref{sid1} against test functions it can be shown that $\dbar g_{\Delta}=0$ as a distribution on $Q$ and that furthermore $\dbar g_{\Delta}=(Z\cdot W) f$ when extended off the quadric in accordance with \eqref{ipt2}. This representative can be evaluated to a space-time field using the integral formula \eqref{ipt3}:
\begin{multline*}
 |X|^{\Delta}\lim_{\delta W\rightarrow 0} \int\limits_{\CP^{3}_{X}}\D^{3}W\wedge\left.\left(\frac{\dbar g_{\Delta}}{Z\cdot W}\right)\right|_{Z^{A}=X^{AB}(W_{B}+\delta W_{B})} \\
 = \frac{(I\cdot Y)^{\Delta}|X|^{\Delta}}{(X\cdot Y)^{\Delta-1}}\lim_{\delta W\rightarrow 0}\int \frac{s^{\Delta-1}\,\d s}{X^{AB}A_{A}\delta W_{B} + sX^{AB}B_{A}\delta W_{B}} \,\bar{\delta}^{(\Delta-2)}\!\left(s+\frac{X^{AB}A_{A}\delta W_{B}}{X\cdot Y}\right) \\
 = \frac{(I\cdot Y)^{\Delta}|X|^{\Delta}}{(X\cdot Y)^{\Delta}} \lim_{\delta W\rightarrow 0} \left[ \frac{X\cdot Y}{X\cdot Y-X^{AB}B_{A}\delta W_{B}}+ O(\delta W)\right] = \frac{(I\cdot Y)^{\Delta}|X|^{\Delta}}{(X\cdot Y)^{\Delta}}\,,
\end{multline*}
which is again the correct bulk-to-boundary propagator.

\medskip

In space-time, evaluating the quadratic action on bulk-to-boundary propagators gives the AdS two-point function, equal to the two-point function of local operators of conformal dimension $\Delta$ in a CFT living on the boundary. This calculation was one of the first tests of the AdS/CFT correspondence~\cite{Gubser:1998bc,Witten:1998qj}, and consequently gives a important check for the twistor formalism. On-shell, the free twistor action reduces to \eqref{sta2}, now evaluated on
\begin{equation*}
 \int\D^{3}Z\,\D^{3}W\,\bar{\delta}^{\prime}(Z\cdot W)\,\tilde{g}_{\Delta}\wedge\dbar g_{\Delta'}\,,
\end{equation*}
with $\tilde{g}_{\Delta},g_{\Delta'}$ of the form \eqref{btbi} and distinct boundary points. The $\D^{3}Z$ and $\D^{3}W$ integrals in this pairing can be evaluated straightforwardly to give:
\begin{multline}\label{s2pt1}
(I\cdot Y_{1})^{\Delta}\,(I\cdot Y_{2})^{\Delta'}\int \D^{3}Z\,\D^{3}W\,\bar{\delta}^{\prime}(Z\cdot W)\,\frac{s^{\Delta-1}\d s}{(W\cdot A)^{\Delta-1}}\,\bar{\delta}^{3}(Z,A(s))\\
\times t^{\Delta'-1}\d t\; \bar{\delta}^{(\Delta'-2)}(Z\cdot C)\;\bar{\delta}^{3}(W,C(t)) \\
=(I\cdot Y_{1})^{\Delta}\,(I\cdot Y_{2})^{\Delta'} \int s^{\Delta-1}\d s\, t^{\Delta'-1}\d t \, \frac{\bar{\delta}^{\prime}(A(s)\cdot C(t))}{(A\cdot C(t))^{\Delta-1}}\,\bar{\delta}^{(\Delta'-2)}(A(s)\cdot C)\,,
\end{multline}
where $Y_{1}^{AB}=A^{[A}B^{B]}$, $Y_{2 AB}=C_{[A}D_{B]}$, $A(s)=A+sB$, and $C(t)=C+tD$. Note that the expression is projectively well-defined only if the two scaling dimensions are equal, so we set $\Delta=\Delta'$. 

The scaling and distributional properties of the remaining portions of the integrand also the $s$ and $t$ integrals to be performed in a basically algebraic manner. It is straightforward to show that \eqref{s2pt1} is equal to
\begin{multline}\label{s2pt2}
 (I\cdot Y_{1})^{\Delta}\,(I\cdot Y_{2})^{\Delta} \int \frac{s^{\Delta-1}\d s\; t^{\Delta-1}\d t}{(B\cdot C)^{\Delta-1} (A\cdot C(t))^{\Delta-1}} \, \bar{\delta}^{\prime}(A(s)\cdot C(t))\,\bar{\delta}^{(\Delta-2)}\!\left(\frac{A\cdot C}{B\cdot C}+s\right) \\
 =(I\cdot Y_{1})^{\Delta}\,(I\cdot Y_{2})^{\Delta} \int \frac{t^{\Delta-1}\,\d t}{(A\cdot C(t))^{\Delta-1}}\frac{(A\cdot C)^{\Delta-1}}{(B\cdot C)^{\Delta}}\,\bar{\delta}^{(\Delta-1)}\!\left(t\left(A\cdot D-\frac{A\cdot C}{B\cdot C}B\cdot D\right)\right) \\
 =(I\cdot Y_{1})^{\Delta}\,(I\cdot Y_{2})^{\Delta} \int \frac{t^{\Delta-1}\,\d t}{(A\cdot C(t))^{\Delta-1}}\frac{(A\cdot C)^{\Delta-1}}{(A\cdot D B\cdot C-A\cdot C B\cdot D)^{\Delta}}\,\bar{\delta}^{(\Delta-1)}(t) \\
 = \frac{(I\cdot Y_{1})^{\Delta}\,(I\cdot Y_{2})^{\Delta}}{(A\cdot D B\cdot C-A\cdot C B\cdot D)^{\Delta}}=\frac{(I\cdot Y_{1})^{\Delta}\,(I\cdot Y_{2})^{\Delta}}{(Y_{1}\cdot Y_{2})^{\Delta}}\,.
\end{multline}
This is precisely the desired form of the 2-point function for massive scalars in AdS$_5$. Written more compactly,
\be\label{s2pt}
\int\D^{3}Z\,\D^{3}W\,\bar{\delta}^{\prime}(Z\cdot W)\,\tilde{g}_{\Delta}\wedge\dbar g_{\Delta'}= \frac{\delta_{\Delta \Delta'}}{(y_{1}-y_{2})^{2\Delta}}\,,
\ee
where $y_{1}, y_{2}\in S^{4}$ lie on the boundary. As expected, this is the two-point function $\la\cO_{\Delta}(y_1)\,\cO_{\Delta'}(y_2)\ra$ of local operators in any four-dimensional CFT.

\medskip

In the larger context of AdS/CFT the bulk partition function of a scalar field with boundary value $\phi$ is equivalent to a generating functional,
\begin{equation*}
\left\la\exp\left(\int_{S^4}\d^{4}y\,\phi(y)\,\cO_{\Delta}(y)\right)\right\ra_{\mathrm{CFT}_4}\,,
\end{equation*}
where $\cO_{\Delta}$ is a local operator in the dual CFT of conformal dimension $\Delta$. The calculation of \eqref{s2pt} demonstrates that the quadratic portion of this functional can be obtained from the twistor space of the AdS$_5$ bulk. It is interesting to note that (at least in some circumstances) there is also a way to express the generating functional in the twistor space of the \emph{boundary}.

As a concrete example, consider chiral primary operators of $\cN=4$ super-Yang-Mills (SYM) in four-dimensions. The simplest of these is the 1/2-BPS operator taking values in the $[0,\mathbf{2},0]$ representation of $\SU(4)$, which can be extended supersymmetrically to the (chiral part of the) stress tensor multiplet (\textit{c.f.}, \cite{Eden:2010zz}). Using the twistor reformulation of $\cN=4$ SYM~\cite{Boels:2006ir,Adamo:2011pv}, this operator can be written succinctly as~\cite{Chicherin:2014uca}:
\be\label{1/2BPS}
\cO_{ijkl}(y)=\int \d^{0|4}\theta_{ijkl}\,\log\det\left(\dbar+\cA\right)|_{Y}\,,
\ee
where $\cA$ is the $\cN=4$ SYM field multiplet written in the twistor space of $S^4$. The operator $(\dbar+\cA)|_{Y}$ is simply a gauge-covariant derivative operator in this twistor space, restricted to a line $Y=A\wedge B$ inside $\CP^3$. In the AdS/CFT dictionary, this operator is dual to a scalar field $\Phi^{ijkl}$ in AdS$_5$ of scaling dimension $\Delta=2$ corresponding to metric components of type IIB supergravity compactified on $S^5$~\cite{Gunaydin:1984fk,Kim:1985ez}. The pairing between boundary data for the bulk scalar and the local operator can then be written in a manifestly covariant form as
\be\label{bpair1}
\int_{S^4} \frac{\d^{4}A\wedge\d^{4}B}{\mathrm{vol}\;\GL(2,\C)}\,\frac{\d^{0|4}\theta_{ijkl}}{(I\cdot Y)^4}\,\phi^{ijkl}(Y)\, \log\det\left(\dbar+\cA\right)|_{Y}\,.
\ee
In principle, the expectation value of this generating functional can be computed on the boundary in a purely twistorial fashion using the $\cN=4$ SYM twistor action. The pairing \eqref{bpair1} can be modified to accommodate more general composite local operators of $\cN=4$ SYM, which themselves can also be written in twistor space~\cite{Koster:2016ebi,Chicherin:2016soh}. 


\subsection{Spinors}

In contrast to the scalar, the standard space-time action for a free AdS spinor of mass $m$ vanishes on-shell. To obtain non-trivial two-point functions, a boundary term which respects the AdS isometries and does not alter equations of motion must be added to the action~\cite{Henningson:1998cd}. In twistor space, the free action is given by generalizing that of the scalar:
\be\label{fta1}
S[\chi,\tilde{\chi}]=\int\D^{3}Z\wedge\D^{3}W\wedge\left[\bar{\delta}^{\prime\prime}(Z\cdot W)\wedge\tilde{\chi}\wedge\dbar \chi - \bar{\delta}(Z\cdot W)\wedge \psi\wedge\tilde{\chi} + \bar{\delta}(Z\cdot W)\wedge\tilde{\psi}\wedge \chi\right]\,,
\ee
where the variational problem involves the off-shell fields $\chi\in\Omega^{0,2}(Q,\cO(\frac{3}{2}-\Delta,\Delta-\frac{5}{2}))$, $\tilde{\chi}\in\Omega^{0,2}(Q, \cO(\Delta-\frac{5}{2}, \frac{3}{2}-\Delta))$, while $\psi\in H^{0,3}(Q, \cO(-\Delta-\frac{1}{2},\Delta-\frac{9}{2}))$ and $\tilde{\psi}\in H^{0,3}(Q,\cO(\Delta-\frac{9}{2},-\Delta-\frac{1}{2}))$ are treated as fixed `sources.' The equations of motion are easily seen to coincide with \eqref{sipt2} for indirect representatives:
\be\label{fteom1}
\bar{\delta}^{\prime\prime}(Z\cdot W)\,\dbar\chi = \bar{\delta}(Z\cdot W)\,\psi \quad \Leftrightarrow \quad \dbar \chi = (Z\cdot W)^{2}\,\psi \,,
\ee
\begin{equation*}
 \bar{\delta}^{\prime\prime}(Z\cdot W)\,\dbar\tilde{\chi} = \bar{\delta}(Z\cdot W)\,\tilde{\psi} \quad \Leftrightarrow \quad \dbar \tilde{\chi} = (Z\cdot W)^{2}\,\tilde{\psi} \,,
\end{equation*}
and the action evaluated on extrema is non-vanishing:
\be\label{fta2}
\left.S[\chi,\tilde{\chi}]\right|_{\mathrm{on-shell}}=\int\D^{3}Z\wedge\D^{3}W\wedge\bar{\delta}^{\prime\prime}(Z\cdot W)\wedge\tilde{\chi}\wedge\dbar \chi\,.
\ee
As in the case of the scalar, the on-shell sources $\psi,\tilde{\psi}$ should be viewed as arising from a separate variational problem.

\medskip

Bulk-to-boundary propagators for spinor fields with scaling dimension $\Delta$ are given in twistor space by modifying those used for the scalar. In particular, a spinor bulk-to-boundary propagator is a solution to the free equation of motion, $K^{\bullet}_{\Delta\,A}(r,x;y)$, where $\bullet$ stands for a boundary Weyl spinor index (\textit{i.e.}, a dotted or un-dotted $\SL(2,\C)$ index). The boundary spinor structure is encoded in twistor space by using the boundary infinity twistor \eqref{infinity} or $I_{AB}=\frac{1}{2}\epsilon_{ABCD}I^{CD}$. For instance, a direct representative with a dotted boundary index, say $\dot{\beta}$, reads:
\be\label{fbtb1}
\psi^{\dot{\beta}}_{\Delta}(Z,W)=[AB]^{\Delta-\frac{1}{2}}I^{BC}B_{C}\frac{\bar{\delta}^{3}_{\Delta-\frac{9}{2}}(W,A)}{(Z\cdot B)^{\Delta+\frac{1}{2}}} + [BA]^{\Delta-\frac{1}{2}}I^{BC}A_{C}\frac{\bar{\delta}^{3}_{\Delta-\frac{9}{2}}(W,B)}{(Z\cdot A)^{\Delta+\frac{1}{2}}}
\ee
Feeding this representative into the integral transform \eqref{sdpt1} gives a space-time formula
\begin{equation*}
 |X|^{\Delta+\frac{1}{2}}\frac{(I\cdot Y)^{\Delta-\frac{1}{2}}}{(X\cdot Y)^{\Delta+\frac{1}{2}}} I^{BC}Y_{CA} = \frac{r^{\Delta+\frac{1}{2}}}{(r^2+(x-y)^2)^{\Delta+\frac{1}{2}}} 
 \left(\begin{array}{cc}
        -\delta^{\dot{\beta}}_{\dot{\alpha}} & y^{\dot{\beta}\delta} \\
        0 & 0
       \end{array}\right)\,,
\end{equation*}
where $Y_{AB}=A_{[A}B_{B]}$ is the boundary point. This expression can be made equal to the standard formula for the spinor bulk-to-boundary propagator~\cite{Henningson:1998cd,Mueck:1998iz} after performing normalized gamma matrix contractions. An un-dotted boundary spinor index is given by taking the dual of \eqref{fbtb1} and replacing $I^{AB}$ with $I_{AB}$ in direct analogy to the scalar case.

An indirect representative for the bulk-to-boundary propagator is given by:
\begin{multline}\label{fbtb2}
\chi^{\dot{\beta}}_{\Delta}(Z,W)= I^{BC}\int s^{\Delta-\frac{1}{2}}\d s\left([AB]^{\Delta-\frac{1}{2}} B_{C} \frac{\bar{\delta}^{3}_{\Delta-\frac{5}{2}}(W,A(s))}{(Z\cdot A)^{\Delta-\frac{3}{2}}} \right. \\
\left. +[BA]^{\Delta-\frac{1}{2}} A_{C} \frac{\bar{\delta}^{3}_{\Delta-\frac{5}{2}}(W,B(s))}{(Z\cdot B)^{\Delta-\frac{3}{2}}}\right)\,,
\end{multline}
with $A(s)=A+sB$ and $B(s)=B+sA$ parametrizing points on the line spanned by $A,B$. As in the case of the scalar, this representative is not obviously $\dbar$-closed:
\begin{multline}\label{fbtb3}
 \dbar\chi^{\dot{\beta}}_{\Delta}=I^{BC}\int s^{\Delta-\frac{1}{2}}\d s\left([AB]^{\Delta-\frac{1}{2}} B_{C}\, \bar{\delta}^{(\Delta-\frac{5}{2})}(Z\cdot A)\,\bar{\delta}^{3}_{\Delta-\frac{5}{2}}(W,A(s)) \right. \\
\left. +[BA]^{\Delta-\frac{1}{2}} A_{C}\,\bar{\delta}^{(\Delta-\frac{5}{2})}(Z\cdot B)\, \bar{\delta}^{3}_{\Delta-\frac{5}{2}}(W,B(s))\right)\,.
\end{multline}
Direct calculation nevertheless shows that $\dbar\chi^{\dot{\beta}}_{\Delta}=0$ as a distribution on $Q$, and furthermore that
\begin{multline}\label{fbtb4}
|X|^{\Delta+\frac{1}{2}} \lim_{\delta W\rightarrow 0} \int_{\CP^3_X} \D^{3}W \wedge W_{A} \left.\left(\frac{\dbar \chi^{\dot{\beta}}_{\Delta}}{(Z\cdot W)^2}\right)\right|_{Z^{A}=X^{AB}(W_{B}+\delta W_{B})} \\
= |X|^{\Delta+\frac{1}{2}}\frac{(I\cdot Y)^{\Delta-\frac{1}{2}}}{(X\cdot Y)^{\Delta+\frac{1}{2}}} I^{BC}Y_{CA}\,,
\end{multline}
as desired.

\medskip

To compute the two-point function in twistor space, the on-shell action \eqref{fta2} is evaluated on bulk-to-boundary representatives. It is straightforward to see that the result is only non-vanishing if one of the representatives has a dotted boundary index and the other has an un-dotted boundary index. Since these representatives contain two terms each, the total integrand of \eqref{fta2} will have four terms. One of these is given by
\begin{multline*}
I_{AC} I^{BD}B_{D} D^{C} [AB]^{\Delta+\frac{1}{2}} \la CD\ra^{\Delta+\frac{1}{2}} \int \D^{3}Z\, \D^{3}W\,\bar{\delta}^{\prime\prime}(Z\cdot W)\, s^{\Delta-\frac{1}{2}}\d s\,t^{\Delta-\frac{1}{2}}\d t\, \\
\frac{\bar{\delta}^{3}(Z, C(s))}{(W\cdot C)^{\Delta-\frac{3}{2}}}\,\bar{\delta}^{(\Delta-\frac{5}{2})}(Z\cdot A)\,\bar{\delta}^{3}(W,A(t))\,,
\end{multline*}
where $\la CD\ra=I_{AB}C^{A}D^{B}$. Each of the other three terms takes a similar form. All of the integrals in this expression can be evaluated against the distributional delta functions to give
\begin{multline*}
 I_{AC} I^{BD}B_{D} D^{C} [AB]^{\Delta+\frac{1}{2}} \la CD\ra^{\Delta+\frac{1}{2}} \int s^{\Delta-\frac{1}{2}}\d s\,t^{\Delta-\frac{1}{2}}\d t\,\bar{\delta}^{\prime\prime}(C(s)\cdot A(t))\,\frac{\bar{\delta}^{(\Delta-\frac{5}{2})}(C(s)\cdot A)}{(A(t)\cdot C)^{\Delta-\frac{3}{2}}} \\
 = I_{AC} I^{BD} [AB]^{\Delta+\frac{1}{2}} \la CD\ra^{\Delta+\frac{1}{2}}\frac{B_{D} D^{C}\,(A\cdot C)^{\Delta-\frac{1}{2}}}{(Y_{1}\cdot Y_{2})^{\Delta+\frac{1}{2}}} \int \frac{t^{\Delta-\frac{1}{2}}\d t}{(A(t)\cdot C)^{\Delta-\frac{3}{2}}}\,\bar{\delta}^{(\Delta-\frac{1}{2})}(t) \\
 =I_{AC} I^{BD} [AB]^{\Delta+\frac{1}{2}} \la CD\ra^{\Delta+\frac{1}{2}} \frac{(A\cdot C)\, B_{D} D^{C}}{(Y_{1}\cdot Y_{2})^{\Delta+\frac{1}{2}}}\,,
\end{multline*}
where $Y_{1\,AB}=A_{[A}B_{B]}$ and $Y_{2}^{AB}=C^{[A}D^{B]}$ are the two distinct boundary points.

Upon combining this expression with the results from the three other terms, one obtains
\begin{multline}\label{f2pt}
\int\D^{3}Z\wedge\D^{3}W\wedge\bar{\delta}^{\prime\prime}(Z\cdot W)\wedge\tilde{\chi}^{\alpha}_{\Delta}\wedge\dbar \chi_{\Delta}^{\dot{\beta}} = (I\cdot Y_1)^{\Delta-\frac{1}{2}} (I\cdot Y_2)^{\Delta-\frac{1}{2}} \frac{ I^{BD}Y_{1\,DE}Y_{2}^{EC} I_{AC}}{(Y_{1}\cdot Y_{2})^{\Delta+\frac{1}{2}}} \\
=\frac{(I\cdot Y_1)^{\Delta-\frac{1}{2}} (I\cdot Y_2)^{\Delta-\frac{1}{2}}}{(Y_{1}\cdot Y_{2})^{\Delta+\frac{1}{2}}} \left(
\begin{array}{cc}
 0 & 0 \\
 (y_{1}-y_{2})^{\alpha\dot{\beta}} & 0 
\end{array}\right)\,,
\end{multline}
which in Poincar\'e coordinates is equivalent to the expected two-point function of spinor operators in a four-dimensional CFT:
\be\label{f2pt2}
\left\la j^{\alpha}_{\Delta}(y_1)\,j^{\dot{\beta}}_{\Delta'}(y_2)\right\ra_{\mathrm{CFT}_4}=\delta_{\Delta \Delta'}\,\frac{(y_1 - y_2)^{\alpha\dot{\beta}}}{(y_{1}-y_{2})^{2\Delta+1}}\,.
\ee


\section{Discussion}

In this paper we have investigated the twistor space of AdS$_5$. In particular we constructed explicit twistor representatives for bulk-to-boundary propagators for fields of various spins, and verified that a natural twistor action for these fields reproduces the expected form of the two-point boundary correlation function. 

It is worth noting that the bulk-to-boundary representatives and free twistor actions presented here can be adapted to AdS$_3$ using the language of `minitwistors'~\cite{Hitchin:1982gh,Jones:1984,Jones:1985pla}. The minitwistor space of AdS$_3$ is the quadric $\CP^{1}\times\CP^{1}$ inside $\CP^3$, with space-time points corresponding to conics inside this quadric. Our direct bulk-to-boundary representatives are easily transcribed into the minitwistor Penrose transform, and two-point functions can be obtained analogously.


The situation is somewhat different in AdS$_4$, where the Penrose transform describes only conformally coupled bulk fields. Here, twistor methods have been applied in \cite{Skinner:2013xp,Adamo:2015ina}, with the aim of finding compact expressions for tree-level bulk correlators, but it is not yet clear how to encode the external states in a useful way. We hope that the study of twistor theory and bulk observables in AdS$_5$ may clarify these issues in the AdS$_4$ setting.  

Our further hope is that the results of this paper, in particular the dual role of $Q$ as both the twistor space of  AdS$_5$ and the ambitwistor space of the boundary, can be used to shed light on the AdS/CFT correspondence from a twistor perspective. However, much work remains to be done. The construction of a non-linear theory on $Q$ describing AdS supergravity remains a challenging problem.

\acknowledgments

We thank Rutger Boels, Lionel Mason and Miguel Paulos for very helpful conversations. The work of TA is supported by a Title A Research Fellowship at St. John's College, Cambridge. DS is supported in part by a Marie Curie Career Integration Grant (FP/2007-2013/631289). JW is supported by an STFC studentship. We would also like to thank the Isaac Newton Institute for Mathematical Sciences, Cambridge, for support and hospitality during the programme {\it Gravity, Twistors and Amplitudes} where work on this paper was partly undertaken. This work was supported by EPSRC grant no EP/K032208/1.

\bibliography{AdS5}
\bibliographystyle{JHEP}

\end{document}